\documentclass[showpacs,reprint,twocolumn,floatfix]{revtex4-1}

\usepackage{graphicx,float,amsmath}
\usepackage{natbib}
\usepackage{lipsum}
\usepackage{xcolor}

\begin{document}

\title{Mechanical stress dependence of the Fermi level pinning on an oxidized silicon surface}

\author{H. Li$^1$}
\author{L. Martinelli$^1$}
\author{F. Cadiz$^1$}
\author{A. Bendounan$^2$}
\author{S. Arscott$^3$}
\author{F. Sirotti$^1$}
\author{A.C.H. Rowe$^1$}
\email{alistair.rowe@polytechnique.edu}

\affiliation{$^1$Physique de la Mati\`ere Condens\'ee, Ecole Polytechnique, CNRS, Universit\'e Paris-Saclay, 91128 Palaiseau, France}
\affiliation{$^2$Synchrotron SOLEIL, L'Orme des Merisiers Saint-Aubin, 91192 Gif-sur-Yvette, France}
\affiliation{$^3$Institut d'Electronique, de Micro\'electronique et de Nanotechnologie (IEMN), Universit\'e de Lille, CNRS, Avenue Poincar\'e, Cit\'e Scientifique, 59652 Villeneuve d'Ascq, France}

\begin{abstract}
A combination of micro-Raman spectroscopy and micro-XPS (X-ray photo-electron spectroscopy) mapping on statically deflected p-type silicon cantilevers is used to study the mechanical stress dependence of the Fermi level pinning at an oxidized silicon (001) surface. With uniaxial compressive and tensile stress applied parallel to the $\langle$110$\rangle$ crystal direction, the observations are relevant to the electronic properties of strain-silicon nano-devices with large surface-to-volume ratios such as nanowires and nanomembranes. The surface Fermi level pinning is found to be even in applied stress, a fact that may be related to the symmetry of the Pb$_0$ silicon/oxide interface defects. For stresses up to 240 MPa, an increase in the pinning energy of 0.16 meV/MPa is observed for compressive stress, while for tensile stress it increases by 0.11 meV/MPa. Using the bulk, valence band deformation potentials the reduction in surface band bending in compression (0.09 meV/MPa) and in tension (0.13 meV/MPa) can be estimated.

\end{abstract}
\pacs{}
\maketitle

As silicon devices continue to shrink towards the nanoscale the electronic properties of the silicon/oxide interface, in particular the energy at which the surface Fermi level is pinned, becomes a key factor in the determination of the overall optical and electronic device characteristics \cite{wan2004, shankar2005, demichel2010, lee2012, kibria2014, shenoy2018}. In parallel with reductions in device size, in-built mechanical stress is widely used to improve the \textit{bulk} electronic properties of CMOS devices \cite{hoyt2002}, so the question of its effect on surface Fermi level pinning becomes an important one. While there is some evidence from electrical measurements in flash memories \cite{toda2005}, MOS capacitors \cite{hamada1994, choi2008} and silicon nano-objects \cite{rowe2008, yang2010, yang2011, li2018}, of the way in which stress modifies the surface Fermi level pinning by deep silicon/oxide interface defects, such measurements do not provide direct spectroscopic access to the stress dependence of the pinned Fermi level.

The vast literature on electronic spectroscopy of silicon/oxide interface states using surface sensitive techniques \cite{himpsel1988,yarmoff1992,rochet1997,jolly2001,pierucci2016} only includes a small fraction of works which deal with mechanical-stress related effects. Amongst these, experimental studies tend to deal with the consequences of local, bond length strains on the properties of clean, reconstructed surfaces \cite{landemark1992, koh2003}, while the effect of local strains at the silicon/oxide interface have been studied from a more theoretical perspective \cite{yazyev2006,kovacevic2014}. In parallel with these photo-emission studies, photo-reflectance has also been used to study strain induced shifts in near-surface electronic energy levels \cite{yin1990, daum1993, imai1996}, including a work in which a macroscopic, externally applied stress is used to modify the electronic structure \cite{sohgawa2001}. Although photo-reflectance can be used to estimate surface potential, being an optical technique it is sensitive principally to the mechanical stress dependence of the near-surface, bulk electronic structure. 

Here, the mechanical stress dependence of the pinned Fermi level at a natively-oxidized (001) surface of a statically flexed, silicon cantilever is studied spectroscopically using X-ray photo-electron spectroscopy (XPS) of the Si 2p core levels. Using the known values of the bulk valence band deformation potentials \cite{hensel1963, milne2012} it is then possible to obtain a spectroscopic estimation of the stress-dependence of the surface band bending. 

\begin{figure}[t]
\includegraphics[clip,width=8.5 cm] {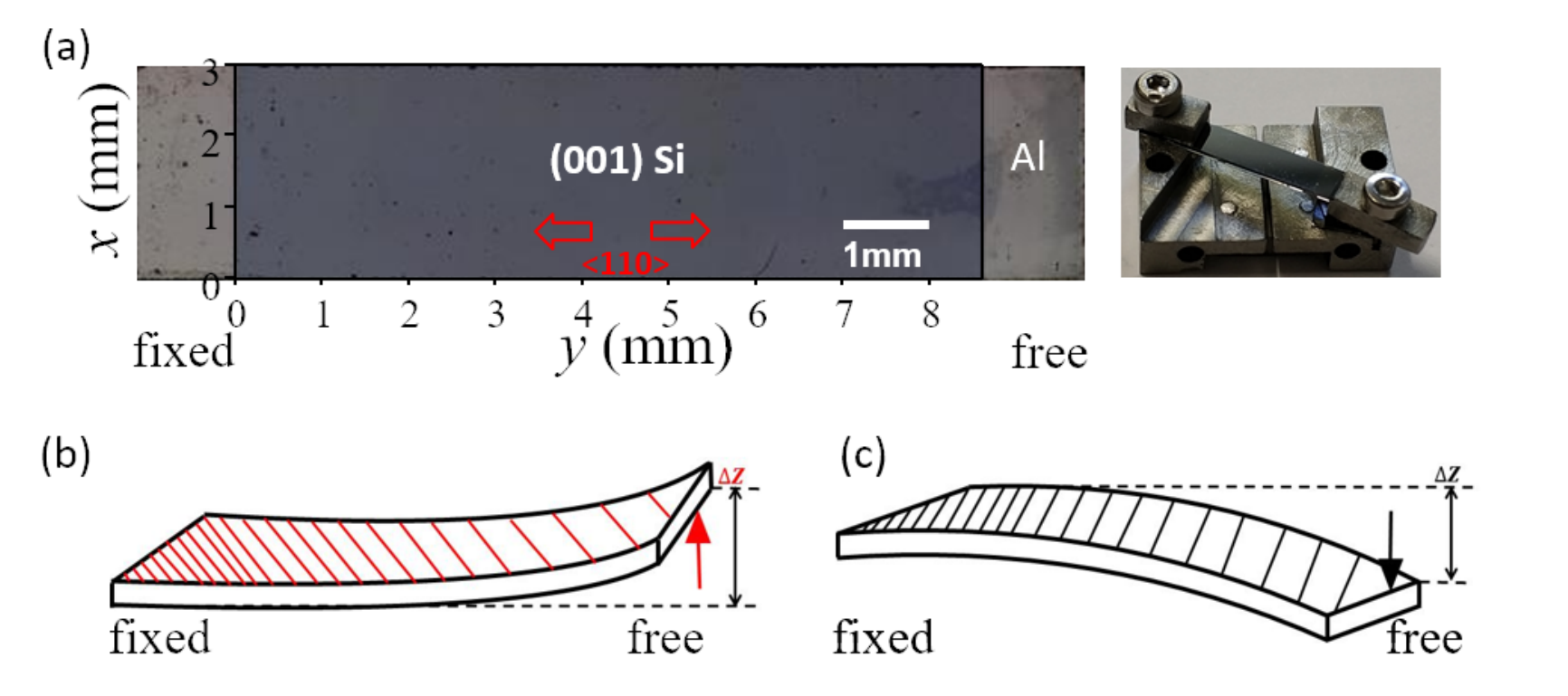}
\caption{(a) p-type silicon cantilevers with ohmic contacts visible at each end that make contact with the earthed sample holder shown right. The \textit{x}- and \textit{y}-axes marked in mm are to locate pixel positions in the subsequent Raman and XPS maps. Using a mechanical screw to apply a force as shown in (b) and (c), the free end of the cantilever is deflected by $\Delta z$. (b) An upwards deflection yields a uniaxial, compressive stress parallel to the $\langle$110$\rangle$ crystal direction on the top surface, with a maximum at the fixed end. (c) A downwards deflection yields an equivalent tensile stress.}
\label{sample}
\end{figure}

Experiments are performed on macroscopic silicon-on-insulator cantilevers fabricated using standard lithographic methods. The boron doped ($\rho <$ 0.01 $\Omega$cm) device layer is 5 $\mu$m thick, the buried oxide layer is 1 $\mu$m thick and the handle is $d = 400$ $\mu$m thick. Once processed (see Supplementary Material), the wafer is diced with a diamond saw into cantilevers $l =$ 11 mm long and whose width, $b =$ 3 mm, with the long axis parallel to the $\langle$110$\rangle$ crystal direction (see Fig. \ref{sample}(a)). Note that the exposed silicon layer which can be accessed for the Raman and XPS mapping is only 8.3 mm long as seen in the photograph in Fig. \ref{sample}(a). The cantilevers are protected during this procedure with a 1 $\mu$m thick photo-resist layer and were subsequently stored with the photo-resist in place for two months. Two weeks prior to the experiments reported here, the photo-resist was removed and the surfaces rinsed in acetone, iso-propyl alcohol and de-ionized water. No further surface treatment was made prior to XPS experiments, a conscience choice made in order to study surfaces more relevant to nano-scale electronic devices than the usual reconstructed surfaces used in XPS studies.

The cantilevers are loaded into sample mounts (seen in Fig. \ref{sample}(a)) with their fixed ends firmly clamped to the mount while the free ends can be deflected upwards or downwards using an actuator screw as shown schematically in Figs. \ref{sample}(b) and (c). This classic cantilevered beam with a free-end point load \cite{hicks2002} results in a uni-axial compressive (tensile) stress parallel to the $\langle$110$\rangle$ crystal direction at the top surface for an upwards (downwards) deflection of the free end. Its magnitude should be maximum at the fixed end and fall linearly to zero at the free end \cite{hicks2002}.  

\begin{figure}[t]
\includegraphics[clip,width=8 cm] {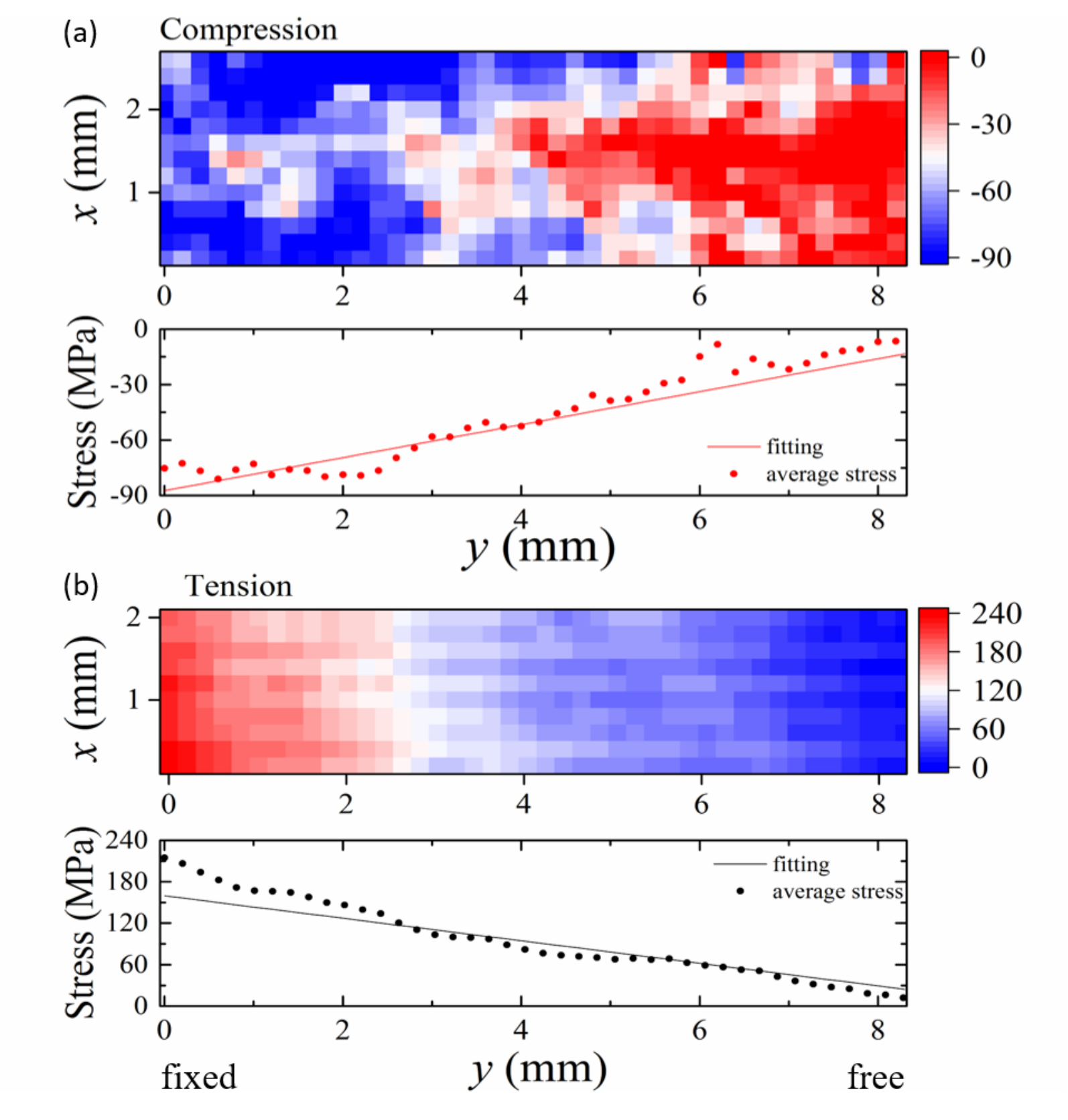}
\caption{Maps of the uniaxial mechanical stress applied parallel to the $y$-direction obtained using micro-Raman spectroscopy for a cantilever whose top surface is under compression (a) and tension (b). The \textit{x-} and \textit{y-}coordinates of the maps correspond to those given in the image of the cantilever in Fig. \ref{sample}(a). The dimensions are smaller than the actual cantilever size, and correspond to the fraction of the surface that was XPS mapped (see Fig. \ref{stress}(a)). Color bars correspond to the measured stress in MPa. The line profiles found by averaging over the measured stress at all \textit{x} values for a given \textit{y} value, agree well with the values found by using the beam formula and the measured amplitude of the free end deflection (see text).}
\label{Raman}
\end{figure}

This is experimentally verified on two statically deflected cantilevers, one with a top surface in compression, one with a top surface in tension, by measuring the sign and magnitude of the stress via the shift in the LO phonon Raman peak \cite{dewolf1996} as a function of position along the length and width of the cantilever. The resulting Raman maps are shown in Figs. \ref{Raman}(a) and (b) respectively. The maps exhibit a quasi-linear increase in the stress from the free to the fixed end, which is particularly clear when the data is averaged over \textit{x-}coordinate values as also shown in Fig. \ref{Raman}. The Raman result can be compared with that expected from textbook formulae by estimating the free end deflection, $\Delta z$, of each cantilever from the measured vertical position of the Raman microscope objective used in autofocus mode. Free end deflections of $\approx 117$ $\mu$m and $\approx -214$ $\mu$m are found in compression and tension respectively. The (maximum) stress at the fixed end is obtained using the formula $X_{\textrm{max}} = 3Ed \Delta z/2l^2$  where $E =$ 170 GPa is Young’s modulus of silicon along the $\langle$110$\rangle$ crystal direction \cite{wortman1965}. Using the cantilever dimensions and free end deflections given above, a maximum mapped stress of approximately -90 MPa and 160 MPa is expected at the fixed end for compression and tension respectively, in excellent agreement with the Raman map data (see solid fitting lines in Fig. \ref{Raman}).

Without any further adjustment of the actuator screws, a spatial map of the kinetic energy, $E_k$, of electrons photo-emitted from the Si 2p core levels is made on each of the two cantilevers. Since the stress is a function of position, a combination of kinetic energy and Raman maps can be used to determine the shift in the Si 2p core level position with mechanical stress $\textit{for a single value of the cantilever deflection, $\Delta z$}$.

\begin{figure}[t]
\includegraphics[clip,width=8.5 cm] {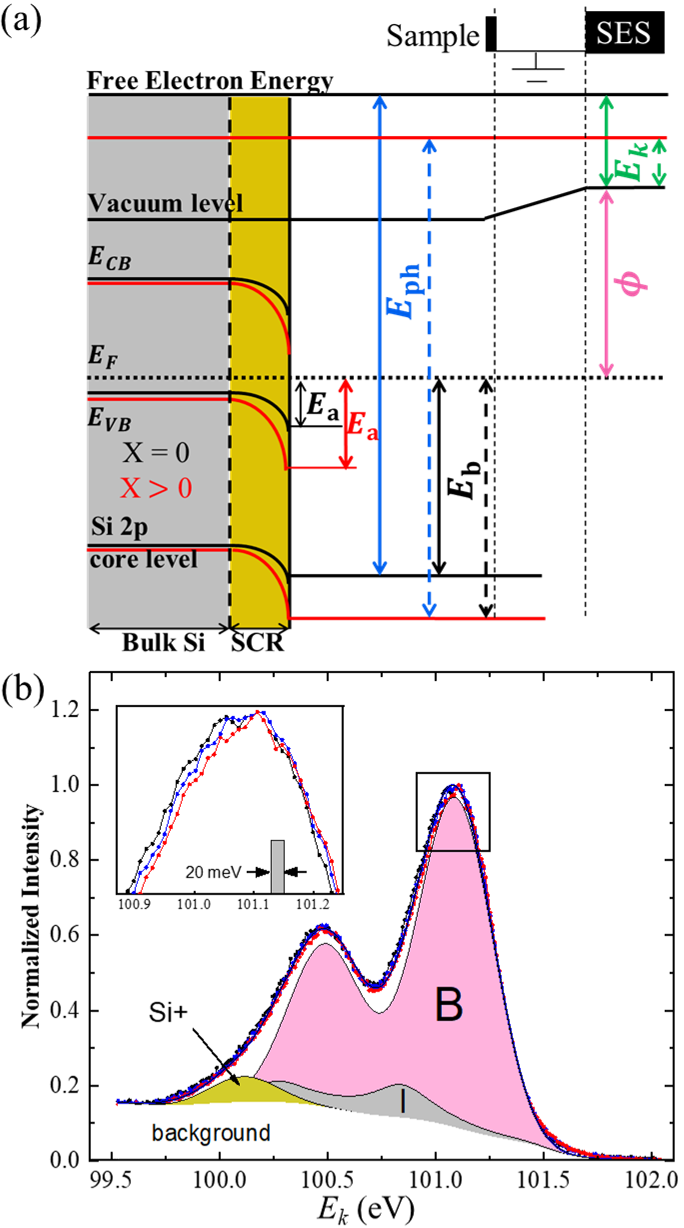}
\caption{(a) Energy level diagram for the surface of p-type silicon showing the electronic structure in the bulk and in the space charge region (SCR) with (red) and without (black) stress. (b) An example of a normalized Si 2p core level XPS spectrum from atoms within 0.12 nm of the silicon/oxide interface for three value of stress, 0 MPa (red), 80 MPa (black) and 160 MPa (blue). Each core level is fitted using Voigt doublets with the parameters shown in Table \ref{fitparams}.}
\label{xps}
\end{figure}

XPS experiments were performed on the ultra-high vacuum (UHV) experimental station of the TEMPO beamline at the Soleil synchrotron using a Scienta SES 2002 electron energy analyzer operated in the swept and fixed modes with a $E_{\textrm{ph}} =$ 200 eV photo-excitation and a 100 eV pass energy. Under these conditions intense Si 2p core level peaks from silicon atoms within a 0.12 nm mean free path \cite{hochella1988, morita1990} of the silicon/silicon oxide interface can be obtained. This extreme interface sensitivity permits a measurement of the stress-dependence of the surface Fermi level pinning energy according to the graphical arguments given in Fig. \ref{xps}(a). The photon energy, $E_{\textrm{ph}}$ and the analyzer work function, $\Phi$, are independent of stress whereas stress modifies the core level binding energy, $E_{\textrm{b}}$, the valence ($E_{VB}$) and conduction band ($E_{CB}$) edges \cite{hensel1963}, and the activation energy of interface defects, $E_a$ \cite{hamada1994} at which the surface Fermi level is pinned. The stress-dependence of the photo-emitted electron kinetic energy, $E_k$ is then equal (but opposite) to $E_{\textrm{a}}$. As indicated by the black ($X = 0$) and red ($X > 0$) in Fig. \ref{xps}(a), knowledge of the stress-dependence of the $E_{VB}$ can then be used to translate this into the stress-dependence of the band bending itself.

Fig. \ref{xps}(b) shows example Si 2p core level spectra plotted as a function of $E_k$ for three different tensile stresses, 0 MPa (red), 80 MPa (blue) and 160 MPa (black). The shift due to the applied stress is clear from the raw data shown (inset). With an SES analyzer energy resolution of about 50 meV, and using the usual Voigt function fitting procedure \cite{landemark1992, gallet2017}, the relative positions of intense Si 2p core level peaks can be determined to within approximately 2 meV (see the Supplementary Material). In the fitting procedure each state is represented by a doublet with a spin-orbit splitting of 0.6 eV, and an intensity ratio 1:2 between 1/2 and 3/2 spin-orbit split states. The intensity, area, kinetic energy shift, and Gaussian broadening of the Voigt functions for the spectrum shown in Fig. \ref{xps}(b) are summarized in Table \ref{fitparams}.

\begin{figure}[t]
\includegraphics[clip,width=8.5 cm] {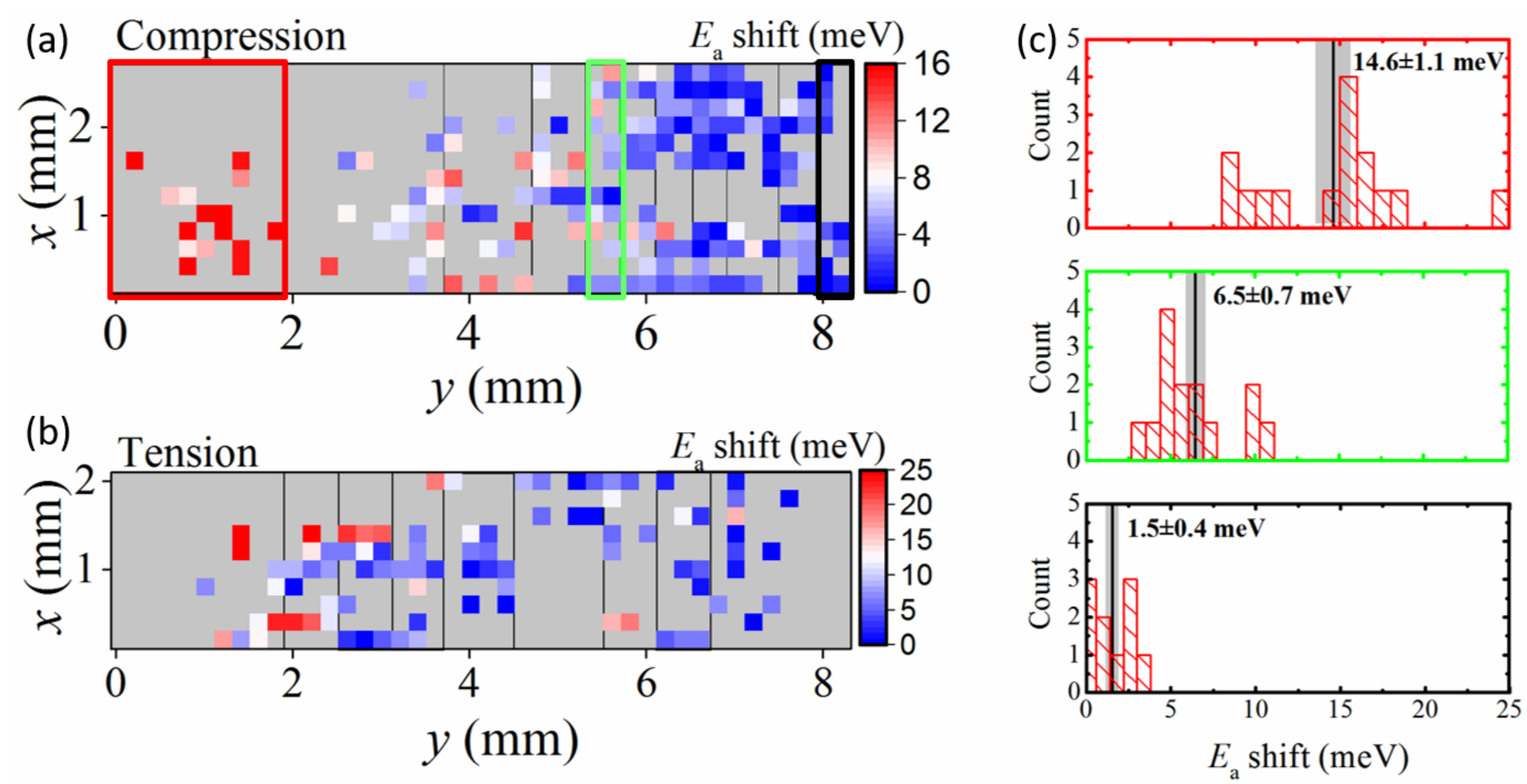}
\caption{Maps of the surface Fermi level pinning energy, $E_{\textrm{a}}$, relative to that obtained at the free end of the cantilever as a function of beam position under (a) compressive stress and (b) tensile stress. The \textit{x-} and \textit{y-}coordinates of the maps correspond to those given in the image of the cantilever in Fig. \ref{sample}(a), and to the Raman maps in Fig. \ref{Raman}. The active (colored) pixels shown in maps are selected because they have the same oxide thickness. (c) The histogram plots of the relative $E_{\textrm{a}}$ of the pixels shown in the colored rectangles of (a). The mean and standard error values of these data, represented as a black line and a gray box respectively, are combined with the micro-Raman spectroscopy data to obtain $E_{\textrm{a}}$ as a function of applied stress.}
\label{stress}
\end{figure}

This procedure is carried out for each pixel in the $E_k$ map, obtained by scanning the cantilever in 100 $\mu$m steps under the soft X-ray beam which is focused to a Gaussian spot of half width $\approx$ 100 $\mu$m. In order to correlate the relative core level peak positions to the applied stress an oxide-thickness related correction must first be applied to the data. As seen in the Supplementary Material the oxide thickness is found to be systematically thicker near the edges of the cantilevers, a fact attributed to the 2-month storage time mentioned above. Unfortunately, oxide thickness influences the Si 2p core level binding energies \cite{eickhoff2004} so that in order to extract a stress-dependence of $E_{\textrm{a}}$, the spatial variation of the oxide thickness must be accounted for. This is achieved by extracting (for each value of the coordinate $x$) only pixels whose oxide thickness is the same along the $y$-direction, at least to within an arbitrarily imposed 2 \% variation around some mean value. The pixel color in Fig. \ref{stress}(a) then represents the binding energy relative to the average value obtained in the rectangle at the free (i.e. zero stress) end of the cantilever. The remaining pixels, shown in gray, are no longer used in the following analysis.

\begin{table}[t]
    \caption{The fitting parameters used in Fig. \ref{xps}(b).}
    \centering
    \begin{tabular}{l|c|c|c|c}
      \textbf{Core level} & \textbf{Label} & \textbf{Intensity} & \textbf{Rel. $E_k$ shift} & \textbf{Width}\\
        &  & $\times 10^{5}$ & (eV) & (eV)\\
      \hline
      Si 2p & B & 22.1 & 0 & 0.4\\
      Si 2p 2$^{\textrm{nd}}$ plane & I & 1.2 & -0.25 & 0.35\\
      Si+ & Si+ & 2.08 & -0.99 & 0.38\\
    \end{tabular}
  \label{fitparams}
\end{table}

\begin{figure}[t]
\includegraphics[clip,width=8.5 cm] {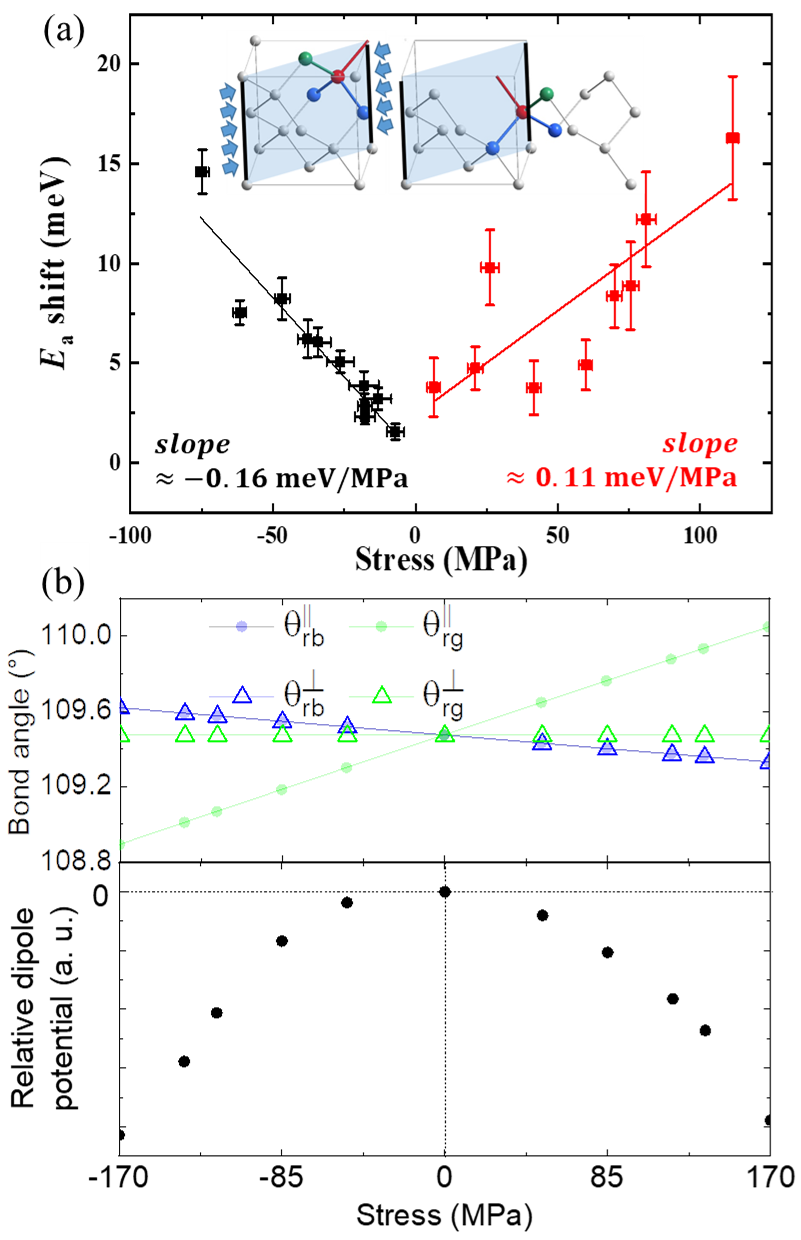}
\caption{(a) The $E_{\textrm{a}}$ shift exhibits an even response in stress. The inset shows the two possible orientations of the Pb$_0$ defect at the (001) surface. Stress is applied in the $\langle$110$\rangle$ direction, and the case of compression is indicated by the blue arrows. The blue $\langle$1$\bar{1}$0$\rangle$ plane serves as a guide to the eye. (b) The stress induced changes in bond angles versus strain, the blue curves correspond to the changes in bond angle between the red dangling bond and the blue back bonds; the green curves correspond to that between the red dangling bond and the green bond. (c) The estimated Coulomb potential energy of an electron in the red dangling bond with respect to stress.}
\label{model}
\end{figure}

In Figs. \ref{stress}(a) and (b), the stress-induced $E_{\textrm{a}}$ shifts, calculated from the kinetic energy of photo-electrons emitted from the Si 2p 3/2 core level according to the schematic in Fig. \ref{xps}(a), are seen to be large and positive around the fixed end ($y =$ 0 mm) of the cantilever for both compressive and tensile stresses, the first suggestion that $E_{\textrm{a}}$ exhibits an even response in applied stress. These two maps are separated into areas containing approximately the same number of pixels (15), denoted by the gray rectangles. The pixels in the red, green and yellow rectangles on the compressively stressed surface are selected to demonstrate how the data is subsequently plotted in Fig. \ref{model}(a). The data histograms from these three rectangles are shown, color coded, in Fig. \ref{stress}(c). The stress-induced shift in the center-of-mass of these histograms is clearly visible, moving from lower $E_{\textrm{a}}$ shifts in the yellow rectangle (free end) to higher $E_{\textrm{a}}$ shifts in the red rectangle (fixed end). In each rectangle the mean and standard error of $E_{\textrm{a}}$ shifts are calculated and marked in Fig. \ref{stress}(c) with a black line and gray box respectively. From the fixed to the free end they are 14.6 $\pm$ 1.1 meV, 6.5 $\pm$ 0.7 meV and 1.5 $\pm$ 0.4 meV respectively. These values, which are relatively robust to changes in the rectangle sizes (i.e. the number of points chosen in each rectangle) establish the mean $E_{\textrm{a}}$ shift as a function of the mean $y$-position of the pixels in each rectangle. By performing a similar procedure using the same pixels in the Raman maps of Fig. \ref{Raman}, it is then possible to obtain a plot, shown in Fig. \ref{model}(a), of the stress-induced $E_{\textrm{a}}$ shift versus the applied stress. 

The result of this procedure, shown in Fig. \ref{model}(a), makes the even response in stress explicitly clear. In compression $E_{\textrm{a}}$ increases by 0.16 meV/MPa while for tensile stresses it increases by 0.11 meV/MPa. These stress-induced shifts can be converted into a stress-dependence of the surface band bending i.e. the difference between the valence band edge in the bulk, and the valence band edge at the silicon/oxide interface shown in Fig. \ref{xps}(a). In order to estimate this it is necessary to account for the stress-induced shifts in the bulk valence bands \cite{hensel1963} which are known from piezoresistance measurements on p-type silicon \cite{milne2012}. Both heavy (HH) and light hole (LH) bands shift to higher energies under compressive stress, with deformation potentials of approximately -3.75 eV and -12.5 eV respectively. Note that the negative sign accounts for the fact that by convention compressive stresses are negative. Using these values, the Young’s modulus of silicon, and the known applied stress levels, the surface band bending reduces by 0.09 meV/MPa in compression, and by 0.13meV/MPa in tension. 

The even symmetry of the $E_{\textrm{a}}$ shift is unusual compared to stress-induced variations in bulk electronic levels which are odd in stress \cite{sohgawa2001}, but it is consistent with interface leakage currents measurements \cite{toda2005,choi2008} and more recently to theoretical estimations of stress-induced energy shifts on other silicon surfaces \cite{kovacevic2014}. The exact origin of this even symmetry is not established, but the symmetry of the intrinsic Pb$_0$ Si/SiO$_2$ was previously speculated to be responsible\cite{choi2008,toda2005}. This interpretation is comforted by the fact that stress-induced shifts in $E_{\textrm{a}}$ of Si/SiO$_2$ interface traps obtained by indirect transport measurements on MOS capacitors are of similar magnitude \cite{hamada1994} to the shifts obtained here.

To explore this idea further, Fig. \ref{model}(b) shows the two possible Pb$_0$ interface defect structures on a (001) silicon surface. In the left structure, referred to as the parallel geometry, the red dangling bond points in the $\langle$111$\rangle$ crystal direction and therefore has a spatial component parallel to the $\langle$110$\rangle$ crystal direction along which stress is applied. In the right panel of Fig. \ref{model}(b), referred to as the perpendicular geometry, the red, dangling bond points in the $\langle \bar{1}11 \rangle$ crystal direction and therefore has a spatial component perpendicular to the $\langle$110$\rangle$ crystal direction. In both cases the stress can be visualized as a force applied uniformly along the thick, black edges of the unit cell parallel to the light blue plane. The arrows in the left panel of Fig. \ref{model}(b) represent the applied force that yields in a uni-axial tensile stress. In the following, a simple relative estimation of the Coulomb potential at the end of the dangling bond due to the electronic charge present in the back bonds (shown in blue and green) will be made as a function of the applied stress. The motivation for this\cite{toda2005} is the notion that this energy will determine $E_{\textrm{a}}$ for Pb$_0$ interface traps. 

Using the compliance tensor for silicon\cite{wortman1965}, the effect of applied stress on the bond angles, and hence on the relative changes in distance between the ends of the bonds can be estimated. This can then be used to estimate the stress-induced changes to the total Coulomb potential at the dangling bond. The calculated changes in the back bond angles are shown in Fig. \ref{model}(c). For a tensile stress the bond angle between the red dangling bond and green back bond in the parallel geometry, $\theta_{\textrm{rg}}^{\parallel}$, increases (see filled, green circles in Fig. \ref{model}(c)), while for the perpendicular geometry,  $\theta_{\textrm{rg}}^{\perp}$ slightly decreases due to Poisson’s effect (see filled, green triangles in Fig. \ref{model}(c)). A compressive stress results in the opposite behavior. The bond angle between the red dangling bond and the blue back bonds in the parallel geometry, $\theta_{\textrm{rb}}^{\parallel}$, decreases under tensile stress due to Poisson’s effect (see filled, blue circles in Fig. \ref{model}(c)), while in the perpendicular geometry, $\theta_{\textrm{rb}}^{\perp}$, decreases (see filled, blue triangles in Fig. \ref{model}(c)). Again, the opposite is true for a compressive stress. A simple geometric calculation, assuming constant bond lengths, then yields a relative estimate of the Coulomb potential as a function of applied stress, and this is found to be even in stress, despite the fact that the bond angles themselves are odd in stress. This is possible because the stress-induced changes to $\theta_{\textrm{rg}}^{\parallel}$ are partially compensated by opposite changes in $\theta_{\textrm{rb}}^{\parallel}$ and $\theta_{\textrm{rb}}^{\perp}$. While this is not proof that the symmetry of the Pb$_0$ centers is the sole origin of the even response in the stress-dependence of $E_{\textrm{a}}$, it is a proof of principle that two odd angular contributions can result in an overall even response. 

The comparison of Raman and XPS maps presented here provides a spectroscopic measurement of the stress-dependence of the surface Fermi level pinning at an oxidized (001) silicon surface. For uniaxial compression along the  $\langle 110 \rangle$ crystal direction the pinning changes by 0.16 meV/MPa, while for tensile stress a value of 0.11 meV/MPa is measured. These quantities have previously only been inferred indirectly from transport measurements, and are important input parameters in the design of any nanoscale strained-silicon electronic device. A simplified analysis of the symmetry of the stress response tentatively suggests that the Fermi level is pinned by Pb$_{0}$ interface states. As such this work provides motivation for a more in-depth theoretical study, for example using ab initio methods \cite{yazyev2006, godet2007,kovacevic2014}, aimed at a quantitative evaluation of the symmetry and magnitude of stress-induced energy shifts of Pb$_{0}$ interface states.

\section{Supplementary Materials}

The Supplementary Material contains full details of the sample fabrication procedure, along with details of the oxide thickness variation across the cantilever surface.

\acknowledgements{This work was partially financed by the French Agence Nationale de la Recherche, contract ANR-17-CE24-0005. The authors thank F. Rochet for useful discussions.}

\bibliographystyle{apsrev}

\bibliography{F:/Publications/References}
\end{document}